\documentclass[aps,reprint,superscriptaddress,longbibliography]{revtex4-1}
\usepackage{graphicx} %
\usepackage{dcolumn} %
\usepackage{bm} %
\usepackage{amsmath}
\usepackage{color}
\draft 

\usepackage{floatrow}
\begin{document}



\title{Angular dependent micro-resonator ESR characterization of a locally doped Gd$^{3+}$:Al$_{2}$O$_{3}$ system}


%
\author{I.~S.~Wisby}
\affiliation{National Physical Laboratory, Hampton Road Teddington, TW11 0LW, UK}
\affiliation{Royal Holloway, University of London, Egham, TW20 0EX, UK}
\author{S.~E.~de Graaf}
\affiliation{National Physical Laboratory, Hampton Road Teddington, TW11 0LW, UK}
\author{R.~Gwilliam}
\affiliation{Advanced Technology Institute, Faculty of Electronics and Physical Sciences, University of Surrey, Guildford, Surrey, GU2 7XH, UK}
\author{A.~Adamyan}
\affiliation{Department of Microtechnology and Nanoscience, MC2, Chalmers University of Technology, SE-41296 Gothenburg, Sweden}
\author{S.~E.~Kubatkin}
\affiliation{Department of Microtechnology and Nanoscience, MC2, Chalmers University of Technology, SE-41296 Gothenburg, Sweden}
\author{P.~J.~Meeson}
\affiliation{Royal Holloway, University of London, Egham, TW20 0EX, UK}
\author{ A.~Ya.~Tzalenchuk}
\affiliation{National Physical Laboratory, Hampton Road Teddington, TW11 0LW, UK}
\affiliation{Royal Holloway, University of London, Egham, TW20 0EX, UK}
\author{T.~Lindstr\"{o}m}
\affiliation{National Physical Laboratory, Hampton Road Teddington, TW11 0LW, UK}

\email[]{ilana.wisby@npl.co.uk}

\date{\today}

\begin{abstract}

 Interfacing rare-earth doped crystals with superconducting circuit architectures provides an attractive platform for quantum memory and transducer devices. Here we present the detailed characterization of such a hybrid system: a locally implanted rare-earth Gd$^{3+}$ in Al$_2$O$_3$ spin system coupled to a superconducting micro-resonator. 
 
 We investigate the properties of the implanted spin system through angular dependent micro-resonator electron spin resonance (micro-ESR) spectroscopy. We find, despite the high energy near-surface implantation, the resulting micro-ESR spectra to be in excellent agreement with the modelled Hamiltonian, supporting the integration of dopant ions into their relevant lattice sites whilst maintaining crystalline symmetries. Furthermore, we observe clear contributions from individual microwave field components of our micro-resonator, emphasising the need for controllable local implantation.

\end{abstract}


\maketitle 

A reliable and scalable quantum information architecture requires a quantum memory \cite{Devoret2013}. A promising route to realizing such a device lies in a hybrid approach, where the unique and desirable properties of a variety of independent physical systems are exploited in conjunction\cite{RevModPhys.85.623}. There has been particular interest in combining superconducting circuits with other two-level systems (TLS) including cold atoms and ions \cite{blinov2004observation, Rosenfeld2007,PhysRevLett.103.043603}, two-level defects \cite{neeley2008process, Falk2013}, molecules \cite{andre2006coherent,Rabl2006} and spin ensembles \cite{Steger2012,PhysRevB.81.241202, Imamoglu2009, Wesenberg2009} - each of which are uniquely suited to individual operation parameters. 
 
One system of particular interest for microwave quantum memory applications are rare earth (RE) ions doped into crystals \cite{Schuster2010,Grezes2014, Kubo2011, Falk2013,Probst2015a,Bienfait2015,Bushev2015}. These ion species are particularly promising as their inner $4f$ optical electronic transitions have very long coherence times \cite{Thiel2011}. Specific RE ions also have the potential for photon conversion between optical and microwave frequency bands for quantum transducer applications \cite{Brien2014,Probst2015a}. 

For now, focus has turned to the requirement for a controllable and scalable infrastructure \cite{Devoret2013}, leading to the development of local ion implantation \cite{Wisby2014,Toyli2010} or specialist focused ion beam \cite{Kukharchyk2014} techniques for precision dopant control which is not otherwise achievable from a growth process alone. With this in mind, we have previously demonstrated a local doping technique utilizing a hard nitride mask to create a locally defined RE spin system, whereby coupling to a superconducting micro-resonator was demonstrated on the order of $3$~MHz \cite{Wisby2014}. While the potential of such a device was evident, it was ascertained from our work, along with that of others \cite{Probst2014}, that when employing such implantation techniques coupling strengths are limited by excessive line-width broadening on the scale of $ \approx 50-100$~MHz.
 
 It is therefore imperative that we study these locally doped crystals and their structure in detail, on the scale at which coupling is mediated. This is, however, challenging due to the nature of local-implantation, where the number of dopant ions is considerably lower (of order $10^{11}$) than in grown, doped crystals. Whilst techniques such as conventional electron spin resonance (ESR), photoluminescence, x-ray photoelectron spectroscopy and secondary ion mass spectroscopy lack the sensitivity required to obtain measurements of such a small spin ensembles, ultra-sensitive microwave spectroscopy at milli-kelvin temperatures has shown more promise \cite{Farr2013,Wisby2014,Toida2015, Bienfait2015}.

In this work, we explore the properties of an implanted gadolinium (Gd$^{3+}$) in Al$_2$O$_3$ spin system through angular dependent micro-resonator ESR (micro-ESR) spectroscopy. We find the measured angular dependent micro-ESR spectra to be in excellent agreement with the modelled Hamiltonian. This supports the conclusion that the dopant ions are well integrated into their relevant lattice sites and that crystalline symmetries are maintained. Furthermore, we observe clear contributions from individual microwave field components of our micro-resonator device, emphasising the need for controllable local implantation.


The data shown in this work is obtained using a sample consisting of seven, frequency multiplexed, inductively coupled, NbN superconducting micro-resonators fabricated atop a systematically implanted Gd$^{3+}$ rare-earth ion ensemble. 

The focus of this work is on a lumped element (LE) resonator of centre frequency $\omega_{r}/2\pi = 3.352$~GHz which sits on a R-cut Al$_2$O$_3$ substrate with locally implanted Gd$^{3+}$ in a $100 \times 250$~$\mu$m area (Fig.~1a). The number of spins beneath the resonator is $ N \approx 10^{11}$. 


The samples are fabricated utilizing a silicon-nitride mask technique which is  detailed extensively in previous work \cite{Wisby2014} and is briefly outlined as follows:

The local implantation process comprises initial deposition of alignment markers which are evaporated atop a commercial R-cut Al$_2$O$_3$ wafer. A SiN mask is next created to act as a stopping barrier for the incident ions during the ion implantation process, and is patterned such that only $\mu$m-size exposed windows will be subject to implantation. The depth profile of the implanted ensemble can be altered by tuning the implantation energy and dose parameters. In our case, $^{160}$Gd$^{3+}$ is implanted at room temperature at a dose of $1 \times 10^{14}$~ions/cm$^{2}$ and at an energy of $900$~keV, giving a concentration profile with peak implantation depth of $170$~nm and full-width-half-maximum $= 77$~nm. 

The contaminated SiN mask is then removed and the implanted substrate annealed at $980^{\circ}$C for $1$~hour in order to remove lattice defects and restore surface crystalline quality. A NbN thin film is then deposited and resonator devices are patterned with standard e-beam lithography techniques whilst ensuring alignment to the implanted regions. 
 
\begin{figure}[t!]
\includegraphics[scale = 0.3]{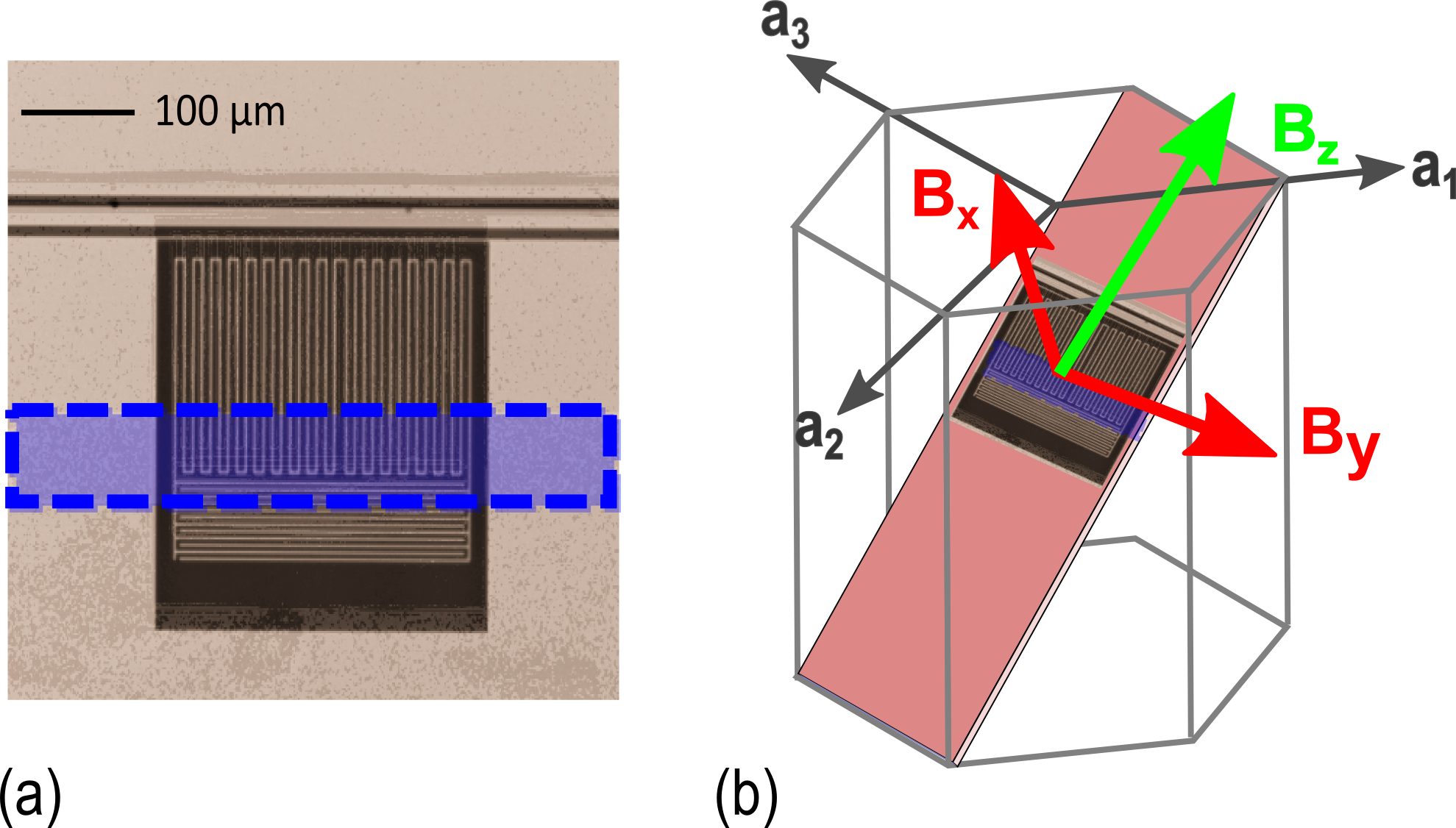}
\caption{(a) An optical image of the LE resonator used in this work. The device is inductively coupled to the transmission line and sits atop a region implanted with Gd$^{3+}$ ions, which is highlighted for clarity with false colouring. (b) The sample is mounted and magnetic field aligned such that the R-plane is parallel to the magnet $B_{z}$ and $B_y$ axes.}
\end{figure}
 

The micro-ESR experiment is performed at mK temperatures in a dilution refrigerator fitted with a vector magnet. The system is equipped with heavily attenuated microwave lines and a low noise cryogenic amplifier.

The magnetic field axis is aligned such that the superconducting plane is parallel to the applied magnetic field $B_{z}$ and $B_{y}$ axes, as shown in Fig.~1b. Initial characterization of the micro-resonator is performed using a vector network analyser (VNA) at  T $\approx 10$~mK, with a power in the resonator of $\approx 3$~pW. From $S_{21}$ measurement we extract internal and coupled quality factors of Q$_{i} = 3.3 \times 10^{5}$, Q$_{c} = 3.8 \times 10^{4}$ respectively - giving a zero field resonator dissipation rate $\kappa / 2 \pi =0.13$~MHz.  

We next perform angular absorption spectroscopy at an enhanced T $= 250$~mK in order to observe higher order transitions. An external static magnetic field ($B_{0}$) is applied in the R-plane of the Al$_2$O$_3$ substrate and is rotated in this plane a full $360^{\circ}$ in $4^{\circ}$ intervals. We assume nominal $0^{\circ}$ rotation where $B_{0} = B_{z}$. For each $B_{0}$ rotation, we apply microwaves on resonance with our micro-resonator and step $B_{0}$ from $0 - 120$~mT, tuning the spin ensemble Zeeman transitions into resonance at spin frequency degeneracies. At each $B_{0}$ step the local $S_{21}$ minima is tracked and $Q_{m} = Q_{i}+Q_{c}$ is extracted the from $S_{21}$ measurements.

\begin{figure}[t!]
\includegraphics[scale = 0.35]{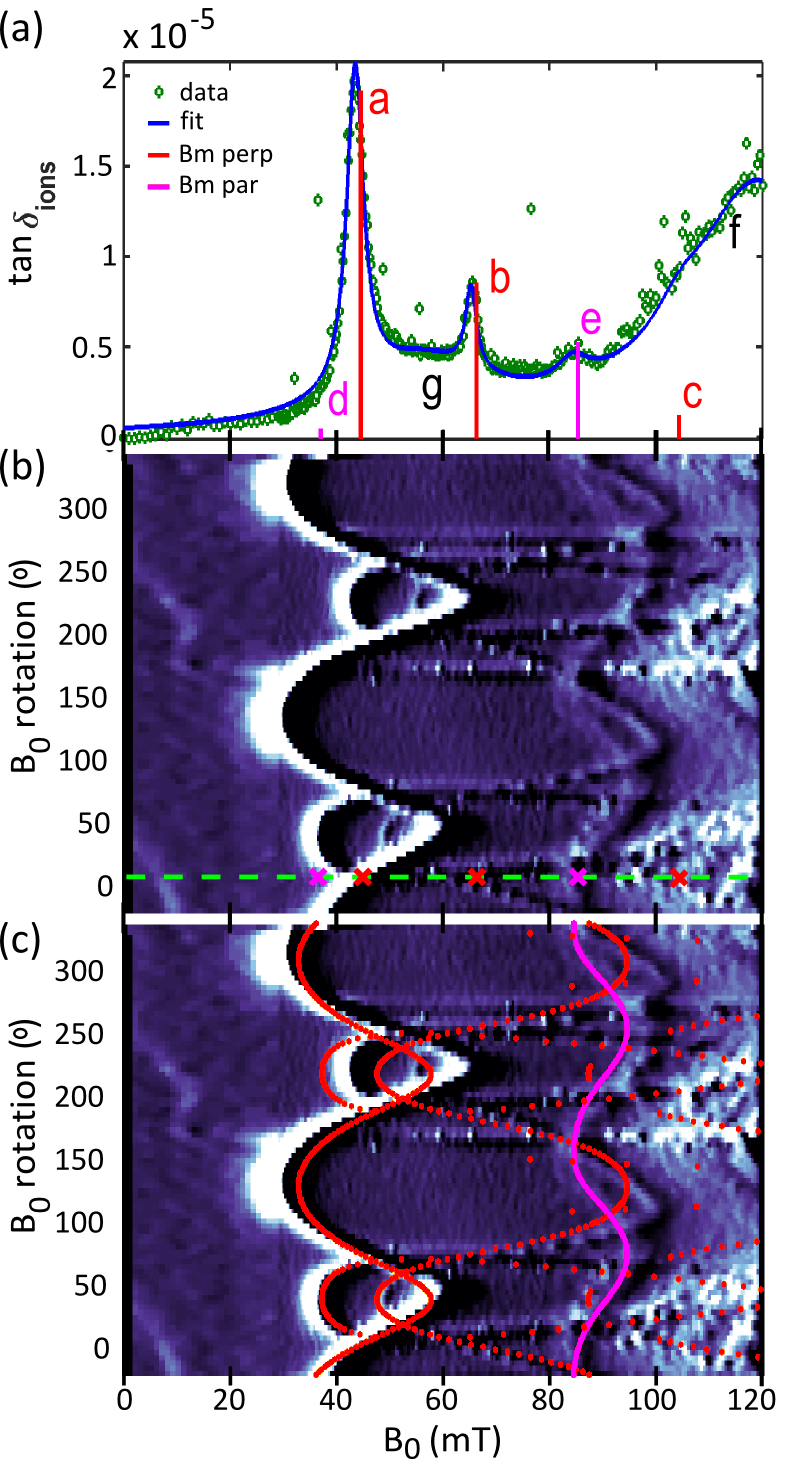}
\caption{(a) An individual absorption spectroscopy trace taken for $B_{0}$ rotated $5^\circ$ (trace position is indicated with green dashed line in Fig.~2b). The resulting $\tan\delta_{ions}$ data is shown (green). Numerical modelling of expected ESR degeneracy points for the Gd$^{3+}$ in Al$_2$O$_3$ spin system in perpendicular mode $B_{m,(a,b,c)} = 44, 66, 104$~mT (red) and parallel mode $B_{m,(d,e)} =37,85$~mT (pink) are marked with respect to relative amplitude obtained from modelling. The fitted overlay (blue) comprises the summation of Gd$^{3+}$:Al$_2$O$_3$ spin system transitions, impurities and an additional unknown transition. Parameters are detailed in Table 2. (b) The total angular dependent ESR spectra intensity plot. An edge detection filter is used to enhance weaker transitions: high contrast correspond to ESR's. (c) The angular ESR spectra with expected peak positions calculated using EASYSPIN numerical modelling. Operation in perpendicular mode (red) and parallel mode (pink).}
\end{figure}

The residual loss tangent due to the ions ($\tan\delta_{ions}$) is extracted from each absorption spectroscopy trace via numerical fitting of the total measured loss tangent: $\tan\delta_{m} = 1/Q_{m} = \tan\delta_{c} + \tan\delta_{int}$, where $\tan\delta_{c}$ is due to coupling to the transmission line. $\tan\delta_{int}$ are intrinsic losses attributable to a summation of the dielectric, magnetic field and losses due to the ions. From $\tan\delta_{int}$ we extract the field dependent loss by subtracting the zero-field losses. The remaining $\tan\delta_{B}$ includes loss due to the ions, while other magnetic field induced losses are assumed to be negligible, such that $\tan\delta_{B} = \tan\delta_{ions}$.

An example of a single absorption spectroscopy trace for a $B_{0}$ rotation of $5^\circ$ is shown in Fig.~2a, where ESR's are observed as an additional absorption mechanism for the microwave photons. The total angular dependent ESR spectra is shown in Fig.~2b as an intensity plot, where we have applied an edge detection filter to enhance weaker transitions such that areas of high contrast correspond to ESR's.


 In order to understand our results, we next examine the ESR spectra using the EASYSPIN \cite{Stoll2006} software package to model the angular dependence of the Gd$^{3+}$:Al$_{2}$O$_{3}$ system within our experimental parameters. In order to achieve this we first consider the spin system and experimental parameters independently.
  
The spin system is described by the Hamiltonian
  \begin{multline*}
   \mathcal{H}=  g \mu_{b} {B} \cdot{S} + B_{2}^{0} O_{2}^{0} + B_{4}^{0} O_{4}^{0} + \\B_{6}^{0} O_{6}^{0} +  B_{4}^{3} O_{4}^{3} +  B_{6}^{3} O_{6}^{3} + B_{6}^{6} O_{6}^{6},
   \end{multline*}
   
  where $O_{k}^{q}$ are hermitian spin operators and $B_{k}^{q}$ coefficients are real parameters. It is customary to redefine these $B_{k}^{q}$ operators in the spin Hamiltonian as:
   
   \begin{table}[h!]
   \begin{tabular}{lclclc}
 
   $b_{2}^{0} = 3B_{2}^{0}$ & $b_{4}^{0} = 60B_{4}^{0}$ & $b_{6}^{0} = 1260B_{6}^{0}$\\
   $b_{4}^{3} = 3B_{4}^{3}$ & $b_{6}^{3} = 36B_{6}^{3}$ & $b_{6}^{6} = 1260B_{6}^{6}$.\\
 
   \end{tabular}
   \end{table}
   
These coefficients have been previously determined experimentally through standard ESR measurements on grown Gd$^{3+}$:Al$_2$O$_3$ for comparable concentrations \cite{Geschwind1961,Meijer1971} at $4.2$K. These parameters are used in our Hamiltonian model, with minor ($<3\%$) adaptation of $b_{2}^{0}$ from $+3123$ to $+3153$~MHz for optimal fitting. The parameters used for modelling in this work are shown in Table.~1. 

 \begin{table}[t!]
 \begin{tabular}{lclc}
\hline
\hline
 $b_{2}^{0} = +3153$ & $b_{4}^{3} = 54.9 $\\
  $b_{4}^{0} = 77.9$ &  $b_{6}^{6} = 14.9 $\\
 $b_{6}^{0} = 3.0 $ & $b_{6}^{3} < 3.0 $ \\
 \multicolumn{2}{c}{$g = 1.9912$} \\
\hline
 \end{tabular}
 \caption{Ground state crystal field splitting parameters of Gd$^{3+}$:Al$_2$O$_3$ in MHz, used for modelling.}
 \end{table}

The Gd$^{3+}$ substitutes into the two inequivalent Al sites of the Al$_2$O$_3$ of $C_{3}$ symmetry. Both sites share the same z-axis, but are rotated about this axis by $60^{\circ}$ with respect to each other. The spin system also has a large zero-field splitting parameter $D = b_{2}^{0}$ attributable to the electric field produced by the O-ions surrounding the Al sites. 

\begin{figure}[t!]
\includegraphics[scale = 0.09]{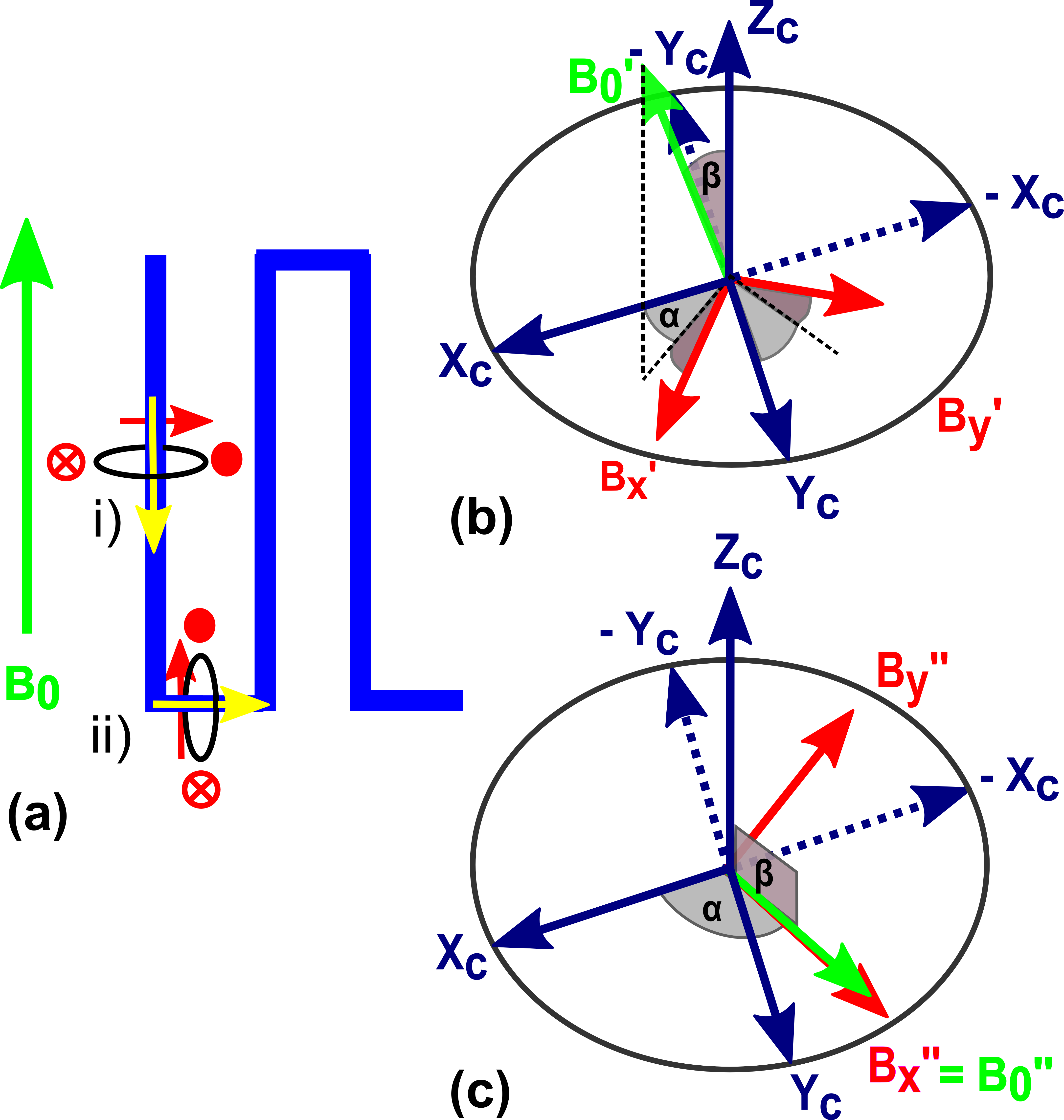}
\caption{\label{mount} (a) With $B_{0}$(green) parallel to nominal magnetic field $B_{z}$ axis the current flowing in the micro-resonator inductor is either parallel (i) or perpendicular (ii) to $B_{0}$. We consider components of the oscillating microwave magnetic field $B_{mw}$ (red) for each case and relative orientation to $B_{0}$.(i) $B_{mw}$ components are perpendicular to $B_{0}$ in all components. (ii) $B_{mw}$ has components both perpendicular and parallel to $B_{0}$. (b) A vector diagram depicting the transformation for case (i) of $B_{0}$ to $B_{0}^{'}$: where $\alpha = 30$ and $\beta = 33$,  $B_{mw}$ components are perpendicular. (c) A vector diagram depicting the transformation for case (ii) of $B_{0}$ to $B_{0}^{''}$: where $\alpha = 120$ and $\beta = 90$. $B_{mw}$ components are perpendicular to $B_{0}$ in $B_{y}^{''}$ and parallel in $B_{x}^{''}$.}
\end{figure}

The experimental system is governed by the geometry of our LE device, which results in  a number of configurations between our static magnetic field $(B_{0})$ and oscillating microwave field component $(B_{mw})$. We begin by considering the case of zero rotation, where $B_{0}$ is parallel to $B_{z}$, such that the current flowing within the micro-resonator inductor is either parallel or perpendicular to $B_{0}$ due to the LE geometry as depicted in Fig.~3a. When the direction of current flow is parallel to $B_{0}$, the oscillating microwave magnetic field component $B_{mw}$ is always perpendicular to $B_{0}$ (Fig.~3a(i)). Where the direction of current flow is perpendicular to $B_{0}$, the $B_{mw}$ component is either perpendicular or parallel to $B_{0}$ (Fig.~3a(ii)).

In order to model such a system, we must transform our static crystal frame into the laboratory frame, accounting for the R-cut crystal. We therefore consider the relative translations of $B_{0}$ from the static crystal frame $[X_{c}$ $Y_{c}$ $Z_{c}]$ into the laboratory frame assuming a start position where $B_{0} = Z_{c}$. In the case where current flow is parallel to $B_{0}$ (Fig.~3a(i)): $B_{0}$ is brought into laboratory frame by a transformation $[\alpha$ $\beta$ $\gamma] = [30$ $33$ $0]$ as shown in Fig.~3b. The resulting $B_{0}^{'}$ is always perpendicular to the $B_{mw}$ component in either $B_{x}^{'}$ or $B_{y}^{'}$. In the case where current flows perpendicular to $B_{0}$ (Fig.~3a(ii)): $B_{0}$ is brought into laboratory frame by a transformation $[\alpha$ $\beta$ $\gamma] = [120$ $90$ $0]$ as shown in Fig.~3c. The resulting $B_{0}^{''}$ is both perpendicular and, importantly parallel to $B_{mw}^{''}$ components $B_{y}^{''}$ and $B_{x}^{''}$ respectively.

We numerically diagonalize our Hamiltonian for each of these operational modes for each $B_{0}$ rotation and find our model is in excellent agreement with the experimental data (Fig.~2c). We find the observed features of higher intensity within our experimental data correspond to operation where $B_{0}$ and $B_{mw}$ components are perpendicular (red overlay in Fig.~2c). The observed features of lower intensity within our experimental data correspond to a parallel $B_{0}$ and $B_{mw}$ (pink overlay in Fig.~2c). An example of this on a single trace is shown in Fig.~2a, where modelled ESR transitions $B_{m}$ are marked: $B_{m,(a,b,c)} = 44, 66, 104$~mT (red) and $B_{m,(d,e)} = 37,85$~mT (pink), distinguished by perpendicular and parallel mode operation respectively.

We can consider the relative areas within which the oscillating magnetic field is flowing both perpendicular and parallel to $B_{0}$ of $3400:832/2$~$\mu m^{2} \approx 8.2:1$ and compare this to relative intensities of the ESR spectra: $8.75:1$, which is accurate to within $10 \%$. 

The resulting features of a single individual absorption trace can be fit with a convolution of functions where the spin ensemble and cavity are modelled as a single mode harmonic oscillator \cite{Schuster2010}: 
\begin{equation}\label{fit}
Q_{m} = \frac{\Delta^{2} + \gamma^{2}}{2g_c^{2} \gamma + \kappa(\Delta^{2} + \gamma^{2})}\omega_{r},
\end{equation}
where $\Delta$ is the detuning from the fitted ESR transition $B_{f}$, $\kappa$ is the cavity line-width  $= 0.13$~MHz, $\gamma$ is the spin line-width and $g_c$ the collective coupling strength. An example of such a fit is overlaid atop an individual absorption spectroscopy trace in Fig.~2a (blue). 

In order to fit the data, we use the numerically modelled ESR degeneracy centre frequencies ($B_{m}$) as start points and constrain $\kappa$. We also include two additional broad ESR features in our function - the first is potentially attributable to Fe$^{3+}$  impurities in sapphire, which when modelled provides an ESR of $B_{m,f} = 119.9$~mT. The second is an unknown background feature previously observed in literature at around $57$~mT \cite{Farr2013, Wisby2014}. For the fit shown in Fig.~2a we extract the parameters detailed in Table.~2. 
  
  \begin{table}[h!]
   \begin{tabular}{lclclclc}
   \hline
   \hline
   \\
Peak  & $B_{m}$ (mT) & $B_{f}$ (mT) & $\gamma$ (MHz) & $g_{c}$ (MHz) & System \\
a & 44 & 43.4 & 53 & 4.5 &  Gd$^{3+}  \perp$ \\
b & 66 & 65.4 & 34 & 1.8 & Gd$^{3+} \perp$ \\
c & 104 & 104.3 & 150 &  2.1 & Gd$^{3+} \perp$ \\
d & 47 & 37.1 & N/A & N/A & Gd$^{3+} \parallel$ \\
e & 85 & 84.1 & 100 & 1.7 & Gd$^{3+} \parallel $\\
f & 119 & 120.0 & 450 & 3.0 & Impurities: Fe$^{3+}$  \\
g & N/A & 57.0 & 500 & 4.3 & Unknown\\
\\
   \hline
   \end{tabular}
   \caption{Extracted parameters from fitting data with a convolution of Eq. 1's.}
   \end{table}
    
It is not surprising that we do not observe the $B_{m,d} = 37$~mT transition, since the expected relative intensity of this transition at $T = 250$~mK is very low. Neglecting $B_{m,d}$, we calculate the root-mean-square deviation between each $B_{m}$ and $B_{f}$ across the entire ESR spectra and find an average of $3\%$ error. We believe this to be in excellent agreement, suggesting that the ions are successfully implanted and crystal structure retained.

 
Notably, our modelled system is obtained using parameters modified by $<3\%$ from measurements obtained in grown Gd$^{3+}$ in Al$_2$O$_3$ crystals \cite{Geschwind1961}. This suggests good integration of the dopant ions into only the relevant lattice sites and that the expected system symmetries are maintained despite the high energy, near surface implantation. 

In these and previous measurements of implanted rare-earth-ion-superconductor systems, coupling strengths have been limited by excessive line-widths \cite{Wisby2014, Probst2014}. The causes of such line-width broadening can be manifold: indicative of inhomogeneous external fields, imperfections in the crystalline environment, i.e. poor site symmetry due to defects, or excessive spin-spin interactions \cite{Abragam1970}.

As our experimental ESR's are in good agreement with our modelled system based on crystal field parameters of grown Gd$^{3+}$ in Al$_2$O$_3$ at similar concentrations \cite{Geschwind1961}, this suggests that dispersion arising from local defects in the crystalline environment may not be the primary cause of our excessive line-width broadening. Notably, line-width broadening is not observed within flip-chip devices coupled to superconducting micro-resonator devices where the relative distances between the two systems are large \cite{Probst}. Neither is broadening prevalent when coupling to other types of cavities where superconducting structures are not present \cite{PhysRevB.90.100404}.

Instead, one possible explanation could be that local field distortions or crystal strain induced by the close proximity of the superconductor, may be accountable for the line-width broadening. It is interesting to note that we do not observe any angular dependence in the extracted line-widths and that the Gd$^{3+}$ parallel contributions have larger line-widths, suggesting a greater relative inhomogeneity in the oscillating magnetic field distribution due to the inductive meander turns. 

It is also worth considering the impact of the resonator geometry on the resulting ESR spectra. We believe this further highlights the need for controllable local implantation techniques in order to create successful hybrid devices integrating spin ensembles and superconducting structures.
 

In conclusion, we demonstrate characterization of a locally implanted Gd$^{3+}$ in Al$_2$O$_3$ system utilizing angular dependent microwave ESR spectroscopy at milli-kelvin temperatures. We find that, despite the high energy and near surface implantation, the resulting angular-ESR spectra is in excellent agreement with the modelled Hamiltonian parametrized by symmetries obtained from a grown Gd$^{3+}$ in Al$_2$O$_3$ system. These results support the successful integration of the dopant ions into their relevant lattice sites. We also observe clear contributions from individual microwave field components of our micro-resonator, emphasising the need for controllable local implantation when coupling spins to superconducting quantum architectures. We further demonstrate that angular-micro-ESR spectroscopy can provide an excellent means of studying even small numbers of spins, giving valuable feedback towards developing appropriate materials for future rare-earth based quantum information technologies. 


We thank, J. Molloy, Y. Andreev, V. Atuchin, A. Zoladek-Lemanczyk, D. Cox for fruitful discussions and support. This work was supported by the NMS, Swedish Research Council (VR) and Linneqs centre. Access to the IBC was supported by the EC program SPIRIT, contract no: 227012.

\bibliographystyle{ieeetr}
\bibliography{Paper_v5}

\end{document}